\documentclass{aa}  
\usepackage{pictex}
\usepackage{graphicx}
\usepackage{color}
\usepackage{natbib}
\usepackage{subfigure} 
\bibpunct{(}{)}{;}{a}{}{,}
\usepackage{txfonts}

\newcommand{\mr}{\mathrm}

\newcommand{\tn}{\textnormal}
\newcommand{\be}{\begin{equation}}
\newcommand{\ee}{\end{equation}}
\newcommand{\bfmath}[1]{\mbox{\boldmath${\mr #1}$\unboldmath}}
\newcommand{\nn}{\nonumber}
\newcommand{\beq}{\begin{eqnarray}}
\newcommand{\eeq}{\end{eqnarray}}

\begin{document}

   \title{Comparing cosmic shear measures}

   \subtitle{Optimizing the Information content of cosmic shear data vectors}

   \author{Tim Eifler \inst{1},
          Martin Kilbinger \inst{1,2},
	  Peter Schneider \inst{1}}
   \offprints{tim.eifler@astro.uni-bonn.de}

   \institute{1)Argelander-Institut f\"ur Astronomie, Universit\"at Bonn, Auf dem H\"ugel 71, D-53121 Bonn, Germany\\
              2) Institut d'Astrophysique de Paris, 98bis boulevard Arago, F-75014  Paris, France}

   \date{}

 
  \abstract
   {}
{We introduce an optimized data vector of cosmic shear measures ($\vec{\mathcal N}$). This data vector has high information content, is not sensitive against B-mode contamination and only shows small correlation between data points of different angular scales.}
{We show that a data vector of the two-point correlation function (2PCF), hereafter denoted as  $\vec \xi$, in general contains more information on cosmological parameters compared to a data vector of the aperture mass dispersion, hereafter referred to as $\vec{\langle M_{\mr{ap}}^2 \rangle}$. Reason for this is the fact that $\vec{\langle M_{\mr{ap}}^2 \rangle}$ lacks the information of the convergence power spectrum ($\mathcal P_\kappa$) on large angular scales, which is contained in $\vec \xi$. Therefore we create a combined data vector $\vec{\mathcal N}$, which retains the advantages of $\vec{\langle M_{\mr{ap}}^2 \rangle}$ and in addition is also sensitive to the large-scale information of $\mathcal P_\kappa$. We compare the information content of the three data vectors by performing a detailed likelihood analysis and use ray-tracing simulations to derive the covariance matrices. In the last part of the paper we contaminate all data vectors with B-modes on small angular scales and examine their robustness against this contamination.}
{The combined data vector $\vec{\mathcal N}$ strongly improves constraints on cosmological parameters compared to $\vec{\langle M_{\mr{ap}}^2 \rangle}$. Although, in case of a pure E-mode signal the information content of $\vec \xi$ is higher, in the more realistic case where B-modes are present the 2PCF data vector is strongly contaminated and yields biased cosmological parameter estimates. $\vec{\mathcal N}$ shows to be robust against this contamination. Furthermore the individual data points of $\vec{\mathcal N}$ show a much smaller correlation compared to $\vec \xi$, leading to an almost diagonal covariance matrix.}
{}


\maketitle
%

\section{Introduction}
Weak gravitational lensing by the Large-Scale Structure (LSS), called cosmic shear, has become a valuable for cosmology. Since the first detection of cosmic shear in 2000 \citep{bre00,kwl00,wme00,wtk00}, several surveys have been carried out with various depth and width. The latest results show the ability of cosmic shear to constrain cosmological parameters, in particular $\sigma_8$ \citep[e.g.][]{vwmh05,smv06,hmv06,tes06,het07,mas07}. These constraints will improve even more in the near future, when  the VST Kilo-degree survey will cover an area of 1700 deg$^2$ with a depth of 15 galaxies per arc minute$^2$, enabling us to estimate the shear signal with less than 1 \% statistical error. This improvement of measuring cosmic shear should go along with an optimization of the data analysis. It is desirable to extract as much information as possible from the observational data and to derive constraints free of any contamination. Currently, most cosmic shear surveys only consider second-order shear statistics, for which all information is contained in the power spectrum of the convergence ($\mathcal P_\kappa$). Although $\mathcal P_\kappa$ is not directly measureable, it is linearly related to second-order cosmic shear measures (e.g. the two-point correlation function and the aperture mass dispersion), which can be estimated from the distorted ellipticities of the observed galaxies. More precisely, all second-order measures are filtered versions of $\mathcal P_\kappa$ and the corresponding filter functions determine how the information content of $\mathcal P_\kappa$ is sampled. It is the intention of this paper to compare several data vectors of cosmic shear measures and to create an optimal data vector with high information content, largely uncorrelated data points and only little sensitivity to a possible B-mode contamination. We first compare the information content of the two-point correlation function (2PCF) and aperture mass dispersion ($\langle M_{\mr{ap}}^2 \rangle$). We prove a general statement that a data vector consisting of 2PCF data points ($\vec \xi$) always gives tighter constraints on cosmological models compared to a data vector consisting of $\langle M_{\mr{ap}}^2 \rangle$ data points ($\vec{\langle M_{\mr{ap}}^2 \rangle}$) and we confirm this by a likelihood analysis of ray-tracing simulations. This result cannot surprise since the 2PCF integrates over all scales of $\mathcal P_\kappa$ and especially collects information on large angular scales which is not taken into account by the aperture mass dispersion. Nevertheless $\langle M_{\mr{ap}}^2 \rangle$ has important advantages. First, it can be used to separate E-modes and B-modes \citep{crit02,svm02}, more precisely $\langle M_{\mr{ap}}^2 \rangle$ is sensitive to E-modes only. Second, due to its narrow filter function it provides highly localized information on $\mathcal P_\kappa$, implying that two different $\langle M_{\mr{ap}}^2 \rangle$ data points are much less correlated compared to the 2PCF. Third, $\langle M_{\mr{ap}}^2 \rangle$ can be easier extended to higher-order statistics \citep{skl05}. These advantages are valuable and should be maintained, but the information content should be improved. Hence, we extend the $\langle M_{\mr{ap}}^2 \rangle$ data vector by one data point of $\xi_+ (\theta_0)$, which provides the large-scale information of $\mathcal P_\kappa$ and call this new data vector $\vec{\mathcal N}$. We perform a likelihood analysis for $\vec{\mathcal N}$, examine its ability to constrain cosmological parameters and compare it to the two aforementioned data vectors. \\
This paper is organized as follows: Sect. \ref{sec:basics} summarizes the basic theoretical background of 2PCF and $\langle M_{\mr{ap}}^2 \rangle$. Next we compare the information content of these two second-order measures and introduce the improvement to the $\langle M_{\mr{ap}}^2 \rangle$ data vector (Sect. \ref{sec:new}). We perform a detailed likelihood analysis for the three data vectors and  present the results in Sect. \ref{sec:sim} and Sect. \ref{sec:likelihoodanalysis}. In Sect. \ref{sec:Bmodes} we contaminate our shear data vectors with B-modes and again perform the likelihood analysis to investigate how significantly each data vector is influenced. Finally in Sect. \ref{sec:conc} we discuss the results and give our conclusions. One final remark should be made on the notation. $\xi$ and $\langle M_{\mr{ap}}^2 \rangle$ denote theoretical quantities calculated from a given power spectrum, whereas $\hat \xi$ and $\mathcal M$ are estimators obtained by averaging over many data points inside a bin. Vectors and matrices are written in bold font. 

\section{Two-point statistics of cosmic shear}
\label{sec:basics}
In this section we briefly review the basics of two-point statistics, definitions of shear estimators and corresponding covariances, closely following the paper of \cite{svk02}. For more details on these topics the reader is referred to \cite{bas01} or, more recently, \cite{sch04}.

\subsection{Two-point correlation function and aperture mass dispersion }
To measure the shear signal we define $\vec \theta$ as the connecting vector of two points and specify tangential and cross-component of the shear $\gamma$ as
\be
\gamma_\mr t = - \mr{Re} \left( \gamma \mr e^{-2\mr i \varphi} \right) \qquad  \tn{and}  \qquad
\gamma_{\times} = - \mr{Im} \left( \gamma \mr e^{-2\mr i \varphi} \right) \;,
\ee 
where $\varphi$ is the polar angle of $\vec \theta$. The 2PCFs depend only on the absolute value of $\vec \theta$ and are defined as
\be
\label{eq:xifrome}
\xi_{\pm} (\theta)= \langle \gamma_\mr t \gamma_\mr t  \rangle (\theta) \pm \langle \gamma_{\times} \gamma_{\times} \rangle (\theta)\;.
\ee
The observed shear field can be decomposed into a gradient component (called E-mode) and a curl component (B-mode) \citep{crit02,svm02}. B-modes are considered to be a contamination of the pure lensing signal, due to noise or unresolved systematics. The limited validity of the Born approximation \citep{jsw00} or redshift source clustering \citep{svm02} can also create B-modes, although these effects are small. Intrinsic alignment of source galaxies is another possible explanation. Predictions about the impact of this effect differ, anyway it can be overcome when using photometric redshifts \citep{ks03}. For the case of a general shear field consisting of E- and B-modes, the convergence is also complex, $\kappa=\kappa_\mr E+ \mr{i}\, \kappa_\mr B$, and it can be related to the shear \citep{ks93} by
\be
\label{eq:ksinvers}
\kappa_\mr E (\vec \theta) + \mr i \kappa_\mr B (\vec \theta)= \frac{1}{\pi} \int \mr d^2 \theta' \mathcal D^\ast (\vec \theta - \vec \theta') \gamma (\vec \theta') \, ,
\ee
with
\be
\mathcal D^\ast (\vec \theta)=\frac{\theta_2^2 - \theta_1^2 + 2 \mr i \theta_1 \theta_2}{|\vec \theta|^4} \,.
\ee
The power spectra of E-mode and B-mode can be defined \citep{svm02} using the Fourier transform of $\kappa$  
\beq
\langle \hat \kappa_\mr E (\vec \ell) \hat \kappa_\mr E^\ast (\vec \ell')\rangle &=&(2\pi)^2 \delta^{(2)} (\vec \ell -\vec \ell') \, \mathcal P_\mr E (\ell) \,,\\
\langle \hat \kappa_\mr B (\vec \ell) \hat \kappa_\mr B^\ast (\vec \ell')\rangle &=&(2\pi)^2 \delta^{(2)} (\vec \ell -\vec \ell') \, \mathcal P_\mr B (\ell) \,,\\
\langle \hat \kappa_\mr E (\vec \ell) \hat \kappa_\mr B^\ast (\vec \ell')\rangle &=&(2\pi)^2 \delta^{(2)} (\vec \ell -\vec \ell') \, \mathcal P_\mr{EB} (\ell)\,,
\eeq
with $\delta^{(2)}(\vec \ell)$ as the two-dimensional Dirac delta distribution. The cross power spectrum $\mathcal P_\mr{EB}$ is expected to vanish for a statistically parity-invariant shear field. Note that $\mathcal P_\mr E$ can be related to the power spectrum of density fluctuations $\mathcal P_\delta$ via Limber's equation \citep{kai92,kai98} 
\beq
\label{eq:pdeltatopkappa}
\mathcal P_\mr E(\ell)&=& \frac{9H_0^4 \Omega_\mr m^2}{4c^4} \int_0^{w_\mr h} 
\frac{\mr d w}{a^2(w)} \mathcal P_{\delta}
\left(\frac{\ell}{f_K(w)},w \right) \nn \\
&\times& \;\left[ \int_w^{w_{\mr h}} \mr d w' p_{w} (w') \frac{f_K (w'-w)}{f_K (w')}\right]^2\;,
\eeq
with $\ell$ as the Fourier mode on the sky, $w$ denotes the comoving coordinate, $w_\mr h$ the comoving coordinate of the horizon, $f_K(w)$ the comoving angular diameter distance and $p_{w}$ the redshift distribution of source galaxies. The 2PCFs depend on both power spectra, $\mathcal P_\mr E$ and $\mathcal P_\mr B$ 
\beq
\label{eq:xi+}
\xi_+(\theta) &=&\int^{\infty}_0 \frac{\mr d\ell\;\ell}{2\pi} \mr J_0(\ell \theta)\, \,\left[\mathcal P_\mr E(\ell) + \mathcal P_\mr B(\ell)\right]\,,\\
\label{eq:xi-}
\xi_-(\theta)&=&\int^{\infty}_0 \frac{\mr d\ell\;\ell}{2\pi} \mr J_4(\ell \theta)\,\left[\mathcal P_\mr E(\ell) - \mathcal P_\mr B(\ell)\right] \,,
\eeq
with $\mr J_\mr n$ denoting the n-th order Bessel function.\\  Another second-order cosmic shear measure, the aperture mass dispersion, was introduced by \cite{swjk98} and is also related to the power spectrum. In contrast to the 2PCF $\langle M_\mr{ap}^2\rangle$ only depends on the E-mode and $\langle M_{\perp}^2\rangle$ only on the B-mode power spectrum, hence the aperture mass statistics provides a powerful tool to separate E- from B-modes
\beq
\label{eq:mapdisper}
\langle M_\mr{ap}^2\rangle (\theta) &=& \frac{1}{2 \pi} \int^\infty_0 d \ell \;
\ell \; \mathcal P_\mr E(\ell) W_\mr{ap}(\theta \ell) \,,\\
\label{eq:mapdisper2}
\langle M_{\perp}^2\rangle (\theta) &=& \frac{1}{2 \pi} \int^\infty_0 d \ell \;
\ell \; \mathcal P_\mr B(\ell) W_\mr{ap}(\theta \ell) \,,
\eeq
with
\be 
W_\mr{ap}(\theta \ell) = \left( \frac{24 \mr J_4 (\ell \theta)}{(\ell
\theta)^2} \right)^2 \, .
\ee
From (\ref{eq:xi+}), (\ref{eq:xi-}) and (\ref{eq:mapdisper}) we see that the second-order shear measures are filtered versions of $\mathcal P_\mr E$ and $\mathcal P_\mr B$. How the different filter functions influence the information content of the corresponding measures will be examined more closely in Sect. \ref{sec:new}. In practice the aperture mass dispersion is difficult to measure due to gaps and holes in the data field but can be expressed in terms of $\xi_+$ and $\xi_-$ as
\be
  \label{eq:map2ofxi}
  \langle M_\mr{ap}^2\rangle (\theta)  =  \int_0^{2\theta} \frac{\mr d \vartheta \, \vartheta}{2 \, \theta^2} \left[ \xi_+ (\vartheta) T_+ \left( \frac{\vartheta}{\theta} \right)\; + \; \xi_- (\vartheta) T_- \left( \frac{\vartheta}{\theta} \right)\right]. 
\ee
The explicit calculation and the filter functions $T_{\pm}$ are given in \cite{svm02}.

\subsection{Estimators}
\label{sec:shearestimators}

Consider a sample of galaxies with angular positions $\vec \theta_i$. For each pair of galaxies we define the connecting vector $\vec \theta= \vec \theta_i-\vec \theta_j$ and determine tangential and cross-components of the ellipticities ($\epsilon_{\mr t}$ and  $\epsilon_{\times}$) with respect to this connecting vector. From these ellipticities we estimate the 2PCF in logarithmic bins of $\vartheta$ with a logarithmic bin width $\Delta \vartheta$ \citep{svk02}. If the bin width is sufficiently small an unbiased estimator for $\xi_{\pm}(\vartheta)$ is given by 
\be
\label{eq:estimatorxi}
\hat \xi_{\pm}(\vartheta) = \frac{1}{N_\mr p(\vartheta)} \sum_{ij} (\epsilon_{i\mr t}\epsilon_{j\mr t} \pm \epsilon_{i\times} \epsilon_{j\times}) \Delta_{\vartheta} (|\bfmath \theta_i -\bfmath \theta_j|) \,,
\ee
with $N_\mr p(\vartheta) = \sum_{ij} \Delta_{\vartheta}(|\bfmath \theta_i - \bfmath \theta_j |)$ as the number of galaxy pairs inside a bin and $\Delta_{\vartheta}(|\bfmath \theta_i-\bfmath \theta_j|)$ is 1 if $|\bfmath \theta_i - \bfmath \theta_j |$ lies inside bin $\vartheta$, 0 otherwise. An unbiased estimator of $\langle M_\mr{ap}^2 \rangle$ can be calculated from $\hat \xi_{\pm}(\vartheta)$ using (\ref{eq:map2ofxi}),
\be
\label{eq:mapcal2}
\mathcal M(\theta_k) = \sum^I_{i=1} \frac{\Delta \vartheta_i \vartheta_i}{2 \,\theta^2_k}  \left[\hat \xi_+(\vartheta_i)\,  T_+ \left(  \frac{\vartheta_i}{\theta_k} \right) + \hat \xi_-(\vartheta_i)\, T_- \left( \frac{\vartheta_i}{\theta_k} \right) \right] ,
\ee
where $I$ must be chosen such that the upper limit of the $I^{\mr{th}}$ bin equals twice the value of $\theta_k$. 
\subsection{Covariances}
\label{sec:covariances}
Important for characterizing the amount of information of a shear estimator is the corresponding covariance. For the 2PCF it is defined as
\be
\label{eq:covxi}
\tn C_\xi \left( \vartheta_i, \vartheta_j \right) := \left \langle
\left( \xi_{\pm} (\vartheta_i)\, - \,\hat \xi_{\pm} (\vartheta_i)\right) \left( \xi_{\pm}(\vartheta_j)\,-\, \hat \xi_{\pm}(\vartheta_j)\right) \right \rangle.
\ee
Assuming a Gaussian shear field the covariance of the 2PCF can be calculated analytically \citep{svk02,jse07}. As one already sees from (\ref{eq:covxi}) the 2PCF has four different covariances, denoted as $\tn C_{++}$, $\tn C_{+-}$, $\tn C_{-+}$, $\tn C_{--}$. Only three of them are independent since $\tn C_{+-}(\vartheta_i ,\vartheta_j)= \tn C_{-+}(\vartheta_j,\vartheta_i)$. The covariance $\tn C_\mathcal M \,(\theta_k,\theta_l)$ of $\mathcal M$ is defined analogously. Using (\ref{eq:mapcal2}) we can express $\tn C_\mathcal M$ in terms of  $\tn C_\xi$
\beq
\label{eq:covmap}
\tn C_\mathcal M (\theta_k,\theta_l)) &=& \frac{1}{4}\sum^{I}_{i=1} \sum^{J}_{j=1} \frac{\Delta \vartheta_i \Delta \vartheta_j}{\theta^2_k \theta^2_l} \; \vartheta_i \vartheta_j \nn \\
&\times& \left[ \sum_{m,n=+,-}T_m \left( \frac{\vartheta_i}{\theta_k} \right) \; T_n \left( \frac{\vartheta_j}{\theta_l} \right) \mr C_{mn}(\vartheta_i,\vartheta_j) \right] \;.
\eeq
Similar to (\ref{eq:mapcal2}) $I$ ($J$) are chosen such that the upper limit of the $I^{\mr{th}}$ ($J^{\mr{th}}$) bin equals twice of $\theta_k$ ($\theta_l$).

\section{The new data vector $\mathcal N$}
\label{sec:new}
Consider two data vectors, namely  
\be
\label{eq:vectorxi}
\vec \xi = \left( \vec \xi_+ \atop \vec \xi_- \right) \quad \tn{with} \quad
\vec \xi_+= \left( \begin{array}{c}
		\xi_+(\vartheta_1)  \\
		\vdots \\
		\xi_+(\vartheta_m) \\
		\end{array}\right) \;,\; 
		\vec \xi_- = \left( \begin{array}{c}
		\xi_-(\vartheta_1)  \\
		\vdots \\
		\xi_-(\vartheta_m) \\
		\end{array}\right) 
\ee
for the 2PCF and
\be
\label{eq:vectormap}
	\vec{\langle M_\mr{ap}^2\rangle} =\left( \begin{array}{c}
	\langle M_\mr{ap}^2\rangle (\theta_1)  \\
	\vdots \\
	\langle M_\mr{ap}^2\rangle (\theta_n) \\
	\end{array}\right)
\ee
for the aperture mass dispersion. The relation (\ref{eq:mapcal2}) can also be written in terms of data vectors and a $n \times 2m$ \textit{transfer matrix} $\bfmath A$
\be
\label{eq:matrixmapxi}
\vec{\langle M_\mr{ap}^2\rangle} = \underbrace{ \left( \begin{array}{c|c}
	\bfmath A_+&\bfmath A_-\\	
	\end{array} \right)\\}_{ \mbox{\large $\mathbf A$}} \left( \vec \xi_+ \atop \vec \xi_- \right)\;,
\ee
with $\bfmath A_+$ denoting the part of $\bfmath A$ referring to $\vec \xi_+$ and $\bfmath A_-$ denotes the corresponding part referring to $\vec \xi_-$. Eq. (\ref{eq:matrixmapxi}) implies that the information content of $\vec{\langle M_\mr{ap}^2\rangle}$ is less or equal compared to $\vec \xi$. The amount of information can only be equal if and only if the rank of $\bfmath A$ equals the dimension of $\vec \xi$, hence rank $\bfmath A =2m$. We explicitly prove these statements in the Appendix. For the case of $\vec \xi$ and $\vec{\langle M_\mr{ap}^2\rangle}$ $n \leq m$ holds, which can be seen from (\ref{eq:mapcal2}). Therefore the relation (\ref{eq:matrixmapxi}) is not invertible and the information content of $\vec{\langle M^2_\mr{ap} \rangle}$ is smaller compared to $\vec \xi_\pm$. The fact that $\vec \xi_\pm$ contains more information on cosmological parameters can also be explained when looking at the filter functions $\mr J_0$, $\mr J_4$ and $W_\mr{ap}$ relating the corresponding second-order shear measures to the underlying power spectrum. $\xi_+$ probes the power spectrum over a broad range of Fourier modes and collects information also on scales larger than the survey size. In contrast, the aperture mass dispersion provides a highly localized probe of $\mathcal P_\mr E$ and does not contain this large-scale information. Hence, due to the limited field size of a survey the information content of $\vec{\langle M^2_\mr{ap} \rangle}$ is smaller compared to $\vec \xi_\pm$. These considerations lead to the idea to modify $\vec{\langle M^2_\mr{ap} \rangle}$ by adding one data point of $\xi_+ (\theta_0)$. We define the new data vector $\vec{\mathcal N}$ as
\be
\label{eq:combination}
	\vec{\mathcal N}=\left( \begin{array}{c}
	\langle M_\mr{ap}^2\rangle (\theta_1)  \\
	\vdots \\
	\langle M_\mr{ap}^2\rangle (\theta_n) \\
	\xi_+(\theta_0) \\
	\end{array}\right)
\ee
and the corresponding covariance matrix reads
\be
\label{eq:covnewmat}
\bfmath C_\mathcal N = \left( \begin{array}{ccc|c}
		\tn{C}_{\mathcal M_{11}} & \cdots& \tn{C}_{\mathcal M_{1n}} & \tn{C}(\mathcal M_1,\xi_+) \\
		\vdots & \ddots & \vdots & \vdots \\
		\tn{C}_{\mathcal M_{1n}} & \cdots& \tn{C}_{\mathcal M_{nn}} & \tn{C}(\mathcal M_n, \xi_+) \\
		\hline
		\tn{C}(\xi_+,\mathcal M_1) &\cdots &\tn{C}(\xi_+,\mathcal M_n)& \tn{C}(\xi_+,\xi_+)
		\end{array} \right).
\ee
The upper left $n\times n$ matrix is exactly $\bfmath C_\mathcal M$ and the entry for C($\xi_+,\xi_+$) is taken from the corresponding covariance matrix of the correlation function. The cross terms can be calculated using (\ref{eq:mapcal2}) and read
\beq
\label{eq:covnew}
\tn{C}(\mathcal M(\theta_k),\hat \xi_+(\theta_0)) &=& \frac{1}{2} \sum^{I}_{i=1} \; \frac{\Delta \vartheta_i}{\theta_k^2} \; \vartheta_i \left[ T_+ \left( \frac{\vartheta_i}{\theta_k} \right) \tn{C}_{++}(\vartheta_i,\theta_0) \right.\nonumber \\
&\;& \left. +  \; T_- \left( \frac{\vartheta_i}{\theta_k}\right) \tn{C}_{-+}(\vartheta_i, \theta_0) \right]\,.
\eeq

\section{Calculating data vectors and covariances}
\label{sec:sim}

The data vectors $\vec \xi, \vec{\langle M^2_\mr{ap} \rangle}, \vec{\mathcal N} $ are directly calculated from the power spectrum of density fluctuations $\mathcal P_\delta$ using (\ref{eq:pdeltatopkappa}) to obtain $\mathcal P_\mr E$ and then applying either (\ref{eq:xi+}), (\ref{eq:xi-}) or (\ref{eq:mapdisper}) depending on the desired cosmic shear measure. To derive $\mathcal P_\delta$ we assume an initial Harrison-Zeldovich power spectrum ($\mathcal P_{\delta}(k) \propto k^n$ with $n=1$). The transition to todays power spectrum employs the transfer function described in \cite{BBKS86}, and for the calculation of the nonlinear evolution we use the fitting formula of \cite{sm03}. In contrast, the covariances are obtained from ray-tracing simulations. The N-body simulation used for the ray-tracing experiment was carried out by the Virgo Consortium \citep{jen01}; for details of the ray-tracing algorithm see \cite{mhb03}. $\bfmath C_\xi$ is calculated by field-to-field variation of 36 ray-tracing realisations, where each field has a sidelength of 4.27 degrees. The intrinsic ellipticity noise is $\sigma_\epsilon=0.3$ and the number density of source galaxies is given by $n=25/\tn{arcmin}^2$. From $\bfmath C_\xi$ we calculate $\bfmath C_{\mathcal M}$ and $\bfmath C_{\mathcal N}$ according to (\ref{eq:covmap}) and (\ref{eq:covnewmat}). The cosmology of the ray-tracing simulations, i.e. our fiducial cosmological model is a flat $\Lambda$CDM model with $\Omega_\mr{m}=0.3$, $\sigma_8=0.9$, $h = 0.7$ and $\Gamma=0.172$.

\subsection{Difficulties with covariances}
\subsubsection{Underestimation of $\bfmath C_{\mathcal M}$}
\label{sec:mixing}
\begin{figure}[htb]
     \includegraphics[width=6cm, angle=270]{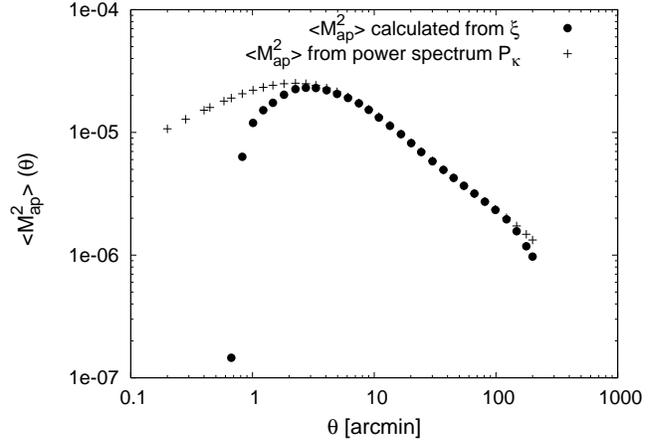}
      \caption{This plot shows $\langle M_{\mr{ap}}^2 \rangle$ calculated directly from the power spectrum (\ref{eq:mapdisper}) compared to $\langle M_{\mr{ap}}^2 \rangle$ calculated from $\xi_\pm$ (\ref{eq:mapcal2}). Due to the fact that we cannot estimate the 2PCF down to arbitrary small angular scales (here $\vartheta_\mr {min}=0 \farcm 2$) the calculated $\langle M_{\mr{ap}}^2 \rangle$ values are underestimated. The same problem occurs when calculating $\bfmath C_\mathcal M$ from $\bfmath C_\xi$. The $\theta$-range with a deviation smaller than 5\% last from $2 \farcm 25 - 100 \farcm 0$.}
         \label{fig:mapcal}
  \end{figure}
\cite{kse06} have shown that $\langle M_{\mr{ap}}^2 \rangle (\theta)$ is biased for small $\theta$ when calculated from the 2PCF using (\ref{eq:mapcal2}). This is due to the lack of 2PCF data points on very small angular scales which causes a small-scale cutoff in the integral of (\ref{eq:map2ofxi}). In our specific case the $\langle M^2_\mr{ap} \rangle$ data vector is not affected by this bias because we calculate it directly from the power spectrum $\mathcal P_\mr E$. However, since $\bfmath C_\mathcal M$ and $\bfmath C_\mathcal N$ are calculated from the covariance of the 2PCF, they are certainly affected by this problem. In this subsection we determine the $\theta$-range on which we can calculate $\bfmath C_\mathcal M$ with sufficient accuracy; the corresponding data vector of the aperture mass dispersion will be restricted to this range. Fig. \ref{fig:mapcal} shows $\langle M_{\mr{ap}}^2 \rangle$ calculated directly from the power spectrum compared with $\langle M_{\mr{ap}}^2 \rangle$ calculated from $\xi_\pm$ using (\ref{eq:mapcal2}). We assume that the deviation shown here is a good approximation for the bias in $\bfmath C_\mathcal M$ and we require an accuracy of 5 \% to accept a $\theta$-value for the $\langle M^2_\mr{ap} \rangle$ data vector. This criterion restricts the data vector to a $\theta$-range of $2 \farcm 25 - 100 \farcm 0$ whereas the 2PCF data vector is measured from $0 \farcm 2 - 200 \farcm 0$.

\subsubsection{Inversion of the covariance matrix}
\label{sec:covinverse}
\begin{figure}[h]
   \includegraphics[width=5.3cm, angle=270]{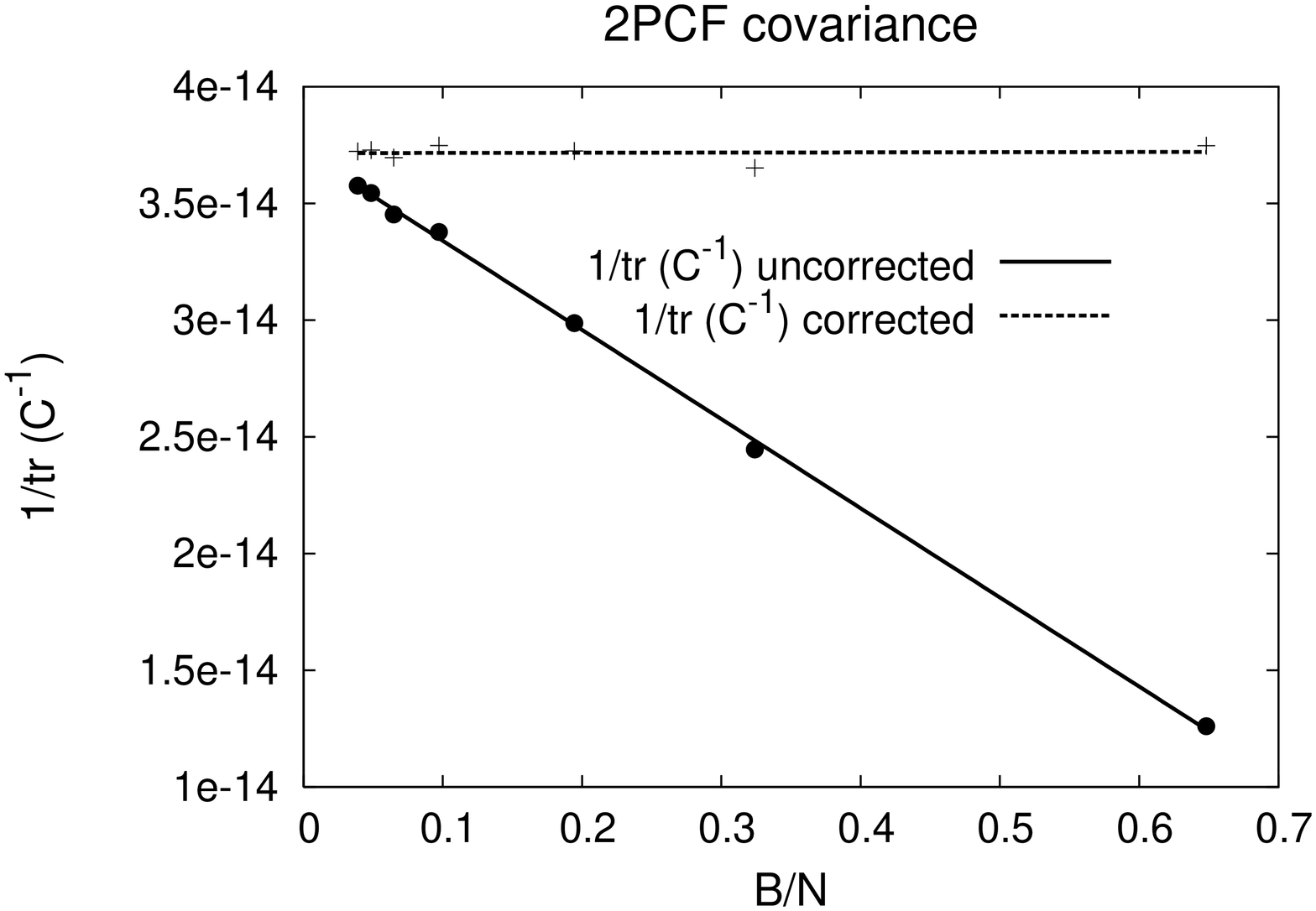}
   \includegraphics[width=5.3cm, angle=270]{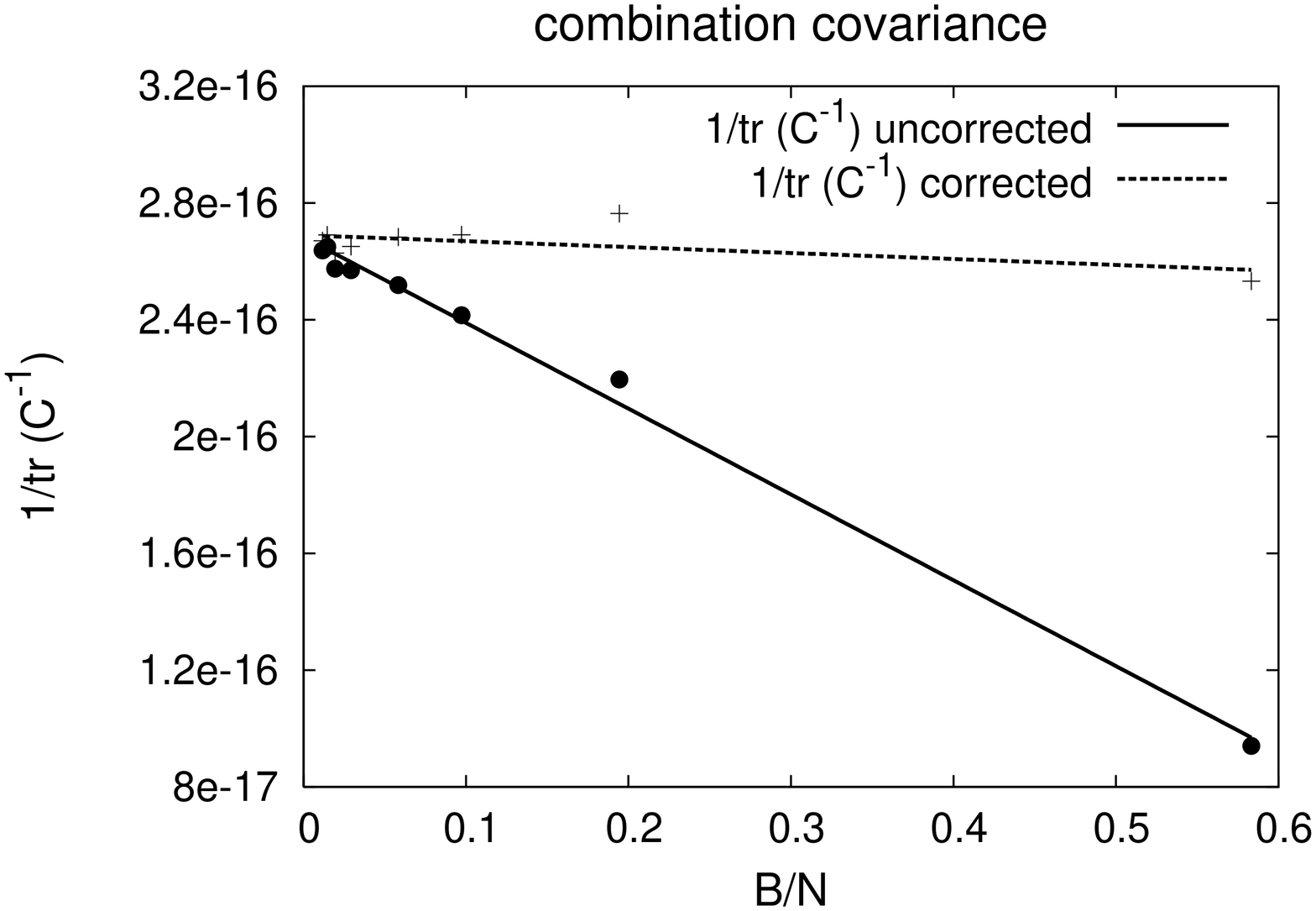}
   \includegraphics[width=5.3cm, angle=270]{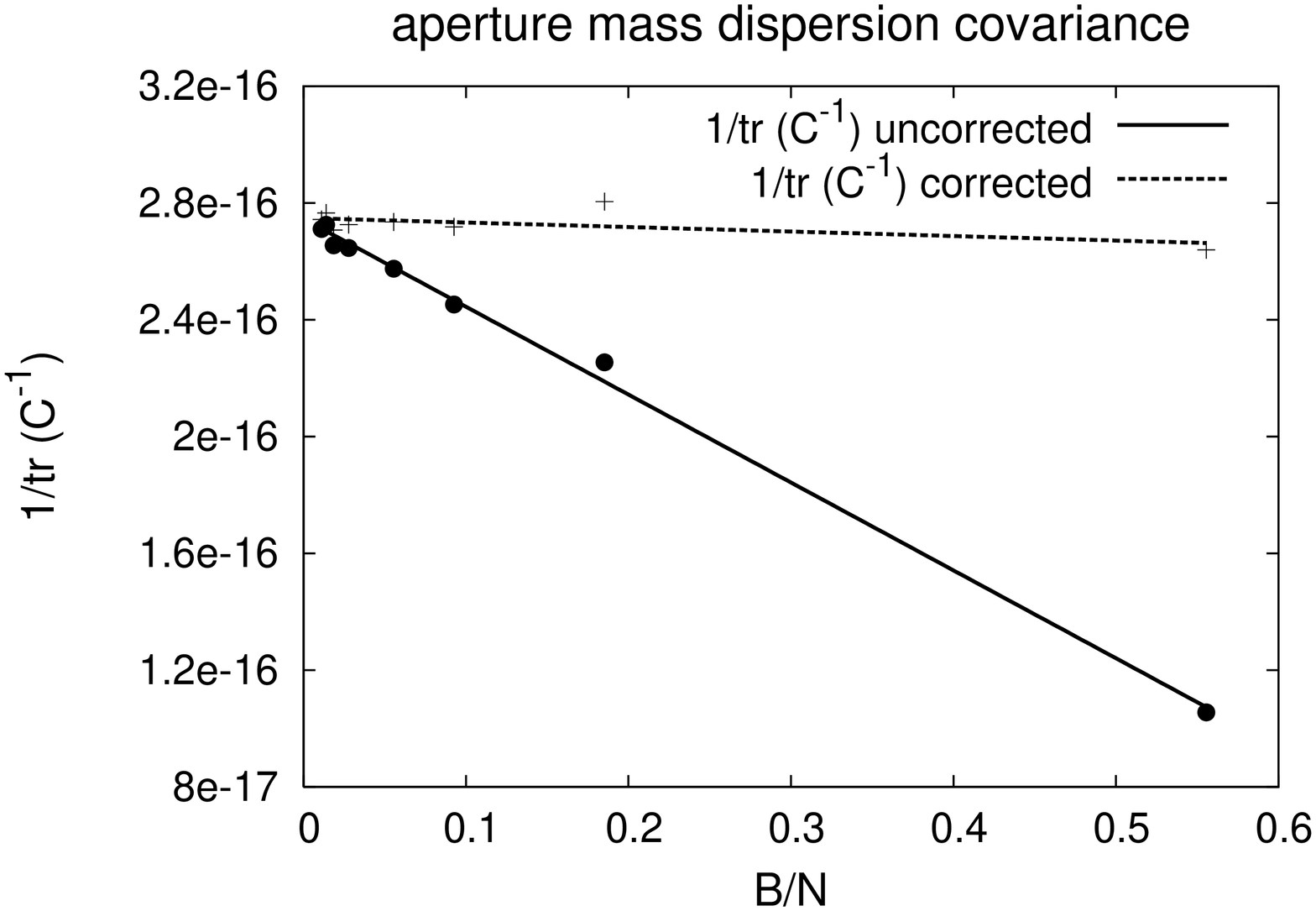}
      \caption{Inverting an estimated covariance matrix yields a bias which depends on the ratio of the number of bins to the number of independent realisations of the ray-tracing simulations ($B/N$). This dependence is linear and we correct the bias for all three inverted covariance matrices $\bfmath C_\xi$, $\bfmath C_{\mathcal N}$, $\bfmath C_{\mathcal M}$. We plot 1/tr ($\bfmath C^{-1}$) for the corrected and uncorrected values where the lines indicate a fit through the data. All covariances are binned logarithmically, $\bfmath C_\xi$ consists of 70 bins covering a range $0 \farcm 2 - 200 \farcm 0$; $\bfmath C_{\mathcal N}$ and $\bfmath C_{\mathcal M}$ cover the range $2 \farcm 25 - 100 \farcm 0$ with 21 bins for $\bfmath C_{\mathcal N}$ and 20 bins for $\bfmath C_{\mathcal M}$.}
         \label{fig:covcheck}
   \end{figure}
A second difficulty in the context of covariance matrices is outlined in \cite{har07}. The fact that an inversion of an estimated unbiased covariance matrix leads to a biased result can be overcome by applying a correction factor. According to \cite{har07} the correction factor depends on the ratio of number of bins $(B)$ to number of independent realisations $(N)$ from which the covariance matrix is estimated. An unbiased estimate of the inverse covariance matrix is
\be
\label{eq:corrfaktor}
\bfmath C^{-1}_{\mr{unbiased}} = \frac{N-B-2}{N-1} \,\bfmath C^{-1} = \left[1- \frac{B+1}{N-1} \right] \,\bfmath C^{-1}.
\ee
\cite{har07} have proven the validity of this correction factor for the case of Gaussian errors and statistically independent data vectors. These two assumptions are violated when estimating the covariance matrix from ray-tracing simulations. In order to check whether the correction factor corrects the error in our ray-tracing covariance matrices, we perform the following experiment. We add different Gaussian noise to the ellipticities of the galaxies, which are taken from the 36 independent realisations of the ray-tracing simulations and thereby increase the number of independent realisations. We hold the binning of the matrices constant, calculate covariances for 36, 108, 216, 360, 720, 1080, 1440, 1800 independent realisations and plot $1/ \tn{tr}$ $\mathbf C^{-1}$ depending on the ratio $B/N$ (Fig. \ref{fig:covcheck}). Note that this method only creates multiple realisations of Gaussian noise on the galaxy ellipticities and does not increase the number of realisations which determine the cosmic variance part of the covariance matrix. Therefore, this method only partly checks for the non-Gaussianity of the errors in a ray-tracing covariance matrix, nevertheless the impact of statistically dependent data vectors is fully taken into account. We find the same linear behavior of the bias as \cite{har07}, therefore we are confident that the correction factor is able to unbias our covariance matrices. Using the corrected inverse covariance matrix we assure that the log-likelihood is also unbiased, nevertheless, any non-linear transformation of the log-likelihood will again introduce a bias which influences the results and must be examined.

\section{Likelihood analysis}
\label{sec:likelihoodanalysis}

We define the posterior likelihood ($P_\mr{PL}$) for the case of a 2PCF data vector as  
\be
P_\mr{PL}(\vec \pi|\vec \xi)=\frac{P_\mr L(\vec \xi|\vec \pi) }{P_\mr E(\vec \xi)} \, P_\mr{Prior}(\vec \pi) \, ,
\ee
where $\vec \pi$ denotes the parameter vector of the $\Lambda \mr{CDM}$ model assumed in our likelihood analysis. $P_\mr{Prior}$ usually contains knowledge on the parameter vector from other experiments. In our case we assume flat priors with cutoffs, which means $P_\mr{Prior}$ is constant for all parameters inside a fixed interval and $P_\mr{Prior}=0$ for parameters outside the interval. The evidence $P_\mr E$, is just the normalization, obtained by integrating the probability over the whole parameter space. The likelihood $P_\mr L$, is defined as 
\be
P(\vec \xi |\vec \pi) = \frac{1}{(2 \pi)^{n/2} \sqrt{\tn{det} \; \bfmath C_\xi}} \exp \left[- \frac{1}{2}\, \chi^2 (\vec \xi, \vec \pi) \right] \, ,
\ee
with the $\chi^2$-function 
\be 
\chi^2 \left(\vec \xi, \vec \pi \right)=(\vec{\xi}(\vec \pi) - \vec{\xi}^\mr f)^\mr t \;\bfmath C^{-1}_\xi \;( \vec{\xi}(\vec \pi) - \vec{\xi}^\mr f) \,.
\ee
$\vec{\xi}^\mr f$ denotes the data vector corresponding to our fiducial model, whereas $\vec{\xi}(\vec \pi)$ varies according to the considered parameter space. To compare the information content of $\vec \xi, \vec{\langle M^2_\mr{ap} \rangle},\vec{\mathcal N} $ we calculate the posterior likelihood in several parameter spaces and illustrate the result by contour plots. Smaller contours correspond to a higher information content.

\subsection{Quadrupole moments}
\label{sec:quadmoments}
\begin{figure}[h]
   \centering
   \includegraphics[width=6cm, angle=270]{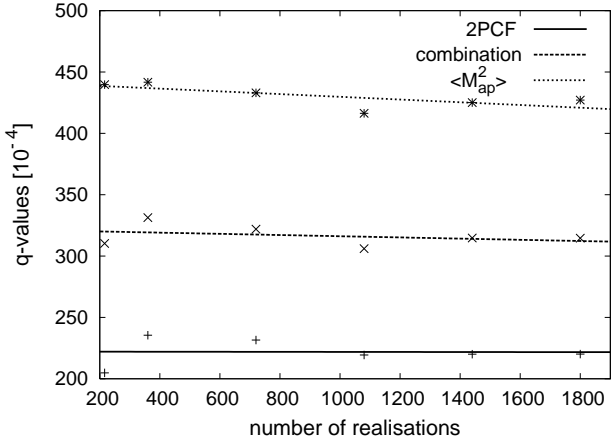}
      \caption{The $q$ of $\vec \xi, \vec{\mathcal N}, \vec{\langle M^2_\mr{ap} \rangle}$ depending on the numbers of independent realisations from which the covariance matrix is estimated. As a parameter space we chose $\Omega_\mr m$ vs. $\sigma_8$. The deviation of $q$ belonging to different numbers of realisations is much smaller than the difference of $q$ of different data vectors.}
         \label{fig:qvalues}
   \end{figure}

In addition to contour plots we illustrate the information content of a data vector by calculating the determinant of the quadrupole moment of the posterior likelihood \citep{kil04}
\be
\label{eq:quadrups}
\mathcal Q_{ij} \equiv \int \tn d^2 \pi \; P_\mr{PL}(\pi_1,\pi_2)(\pi_i-\pi_i^{\mr f})(\pi_j -\pi_j^{\mr f}),
\ee
with $\pi_1$ and $\pi_2$ as the varied parameters, $\pi_i^{\mr f}$ as the parameter of the fiducial model. The calculation of $\mathcal Q_{ij}$ assumes a posterior likelihood in a two-dimensional parameter space, when considering more than two varied parameters we calculate the  $\mathcal Q_{ij}$ for the marginalized posterior likelihood (see Sect. \ref{sec:3para}).
The determinant is given by
\be
\label{eq:detquadrups}
q= \sqrt{\tn{det}\, \mathcal Q_{ij}} = \sqrt{\mathcal Q_{11} \mathcal Q_{22} - \mathcal Q_{12}^2}.
\ee 
Tighter constraints on the parameters correspond to a smaller value of $q$. Due to its non-linearity in the log-likelihood $q$ is biased (Sect. \ref{sec:covinverse}). The amount of bias varies depending on the number of independent realisations from which the covariance matrix is estimated and we examine this effect in a similar way as for the covariance matrices in Sect. \ref{sec:covinverse}. For six different numbers of independent realisations we perform a likelihood analysis in a two-parameter space ($\Omega_\mr m$ vs. $\sigma_8$) and calculate $q$ for all three cosmic shear measures. The result is plotted in Fig. \ref{fig:qvalues}.
One clearly sees that the $q$ dependence on the number of realisations is much weaker compared with the difference between $q$ of different cosmic shear measures. Therefore, the bias is small and we can confidently use $q$ to compare the relative information content of the different data vectors.

\subsection{Variations of two parameters}
\label{sec:2para}
\begin{figure}[h!]
   \centering
   \includegraphics[width=6cm, angle=270]{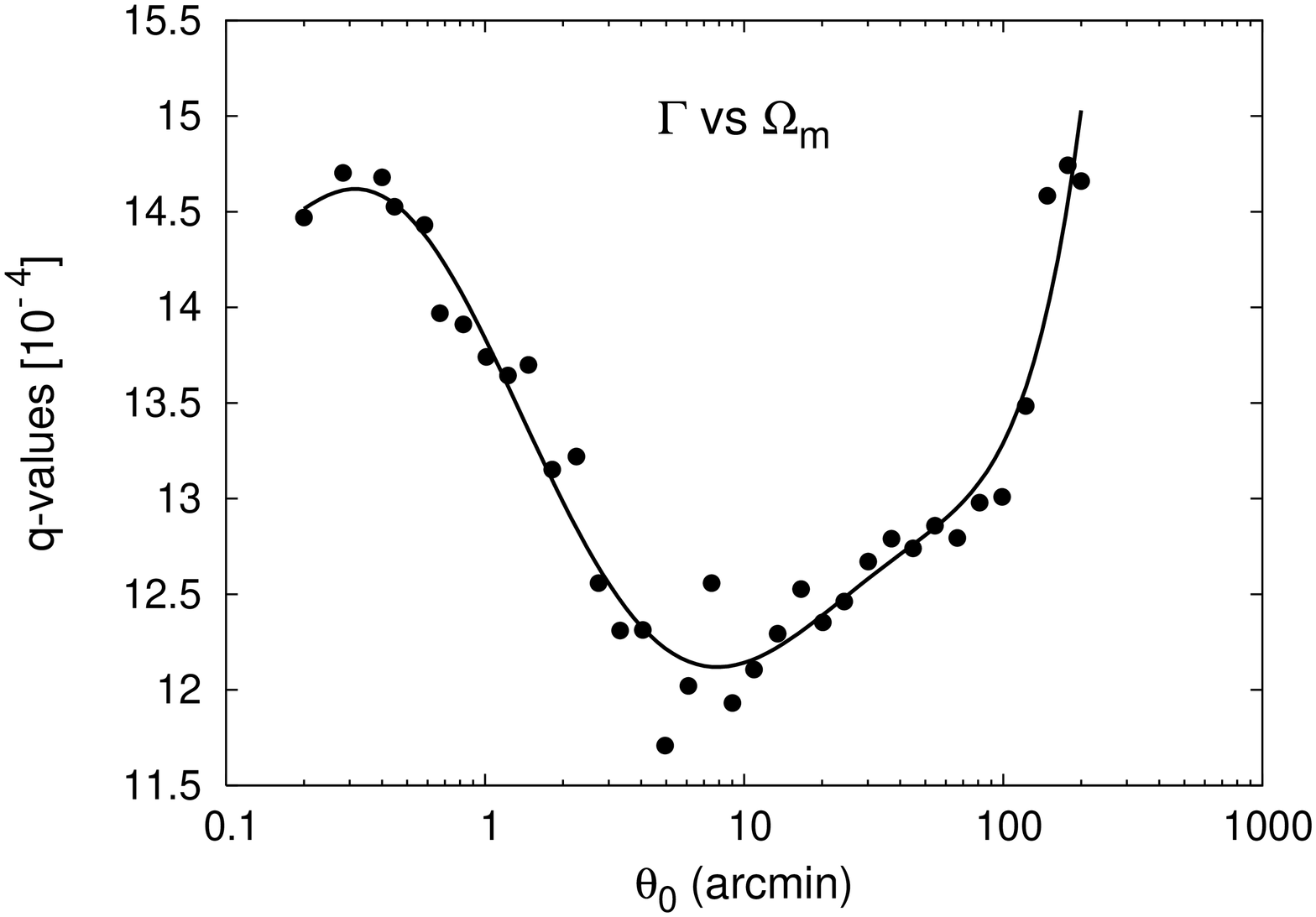}
   \includegraphics[width=6cm, angle=270]{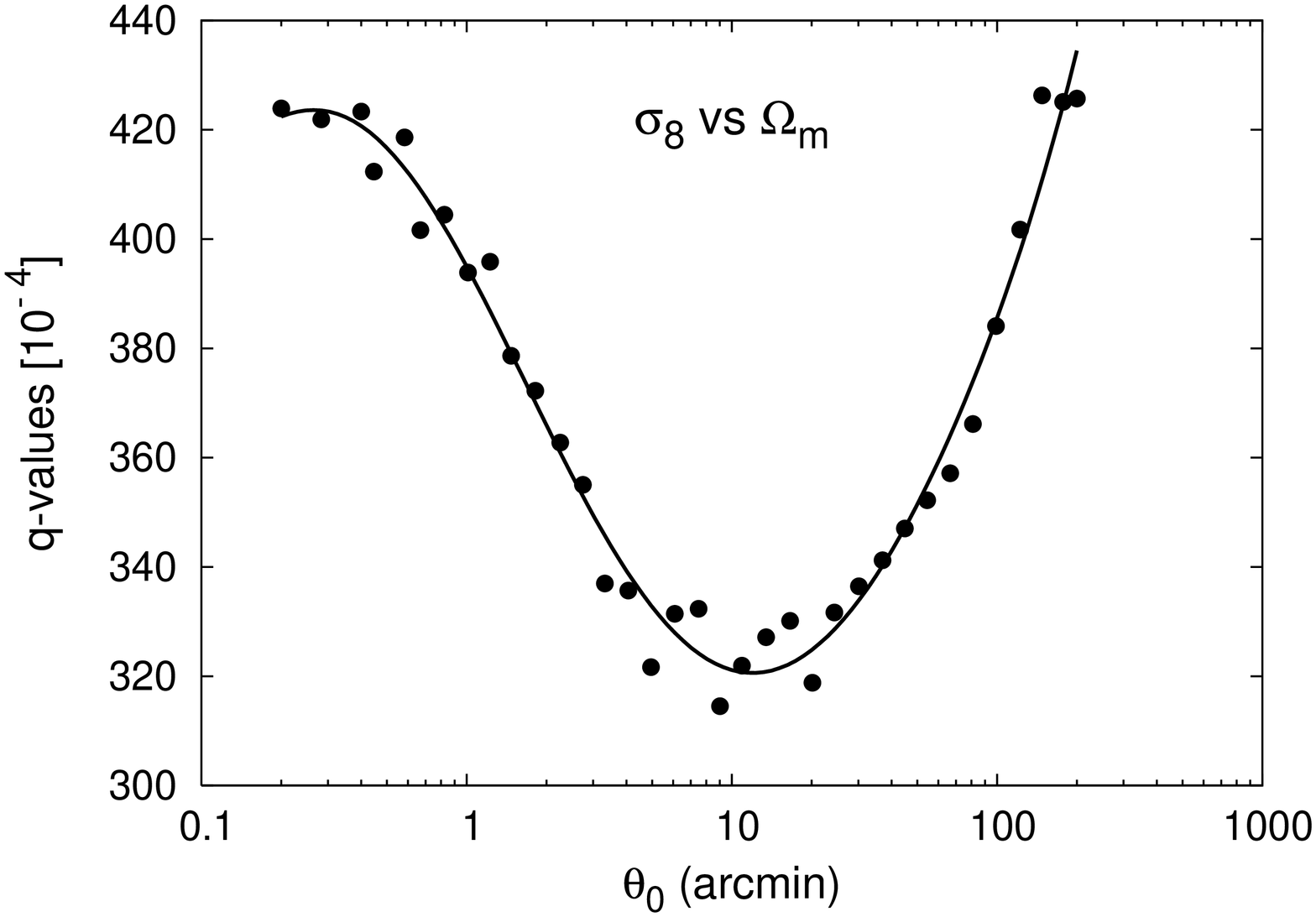}
    \includegraphics[width=6cm, angle=270]{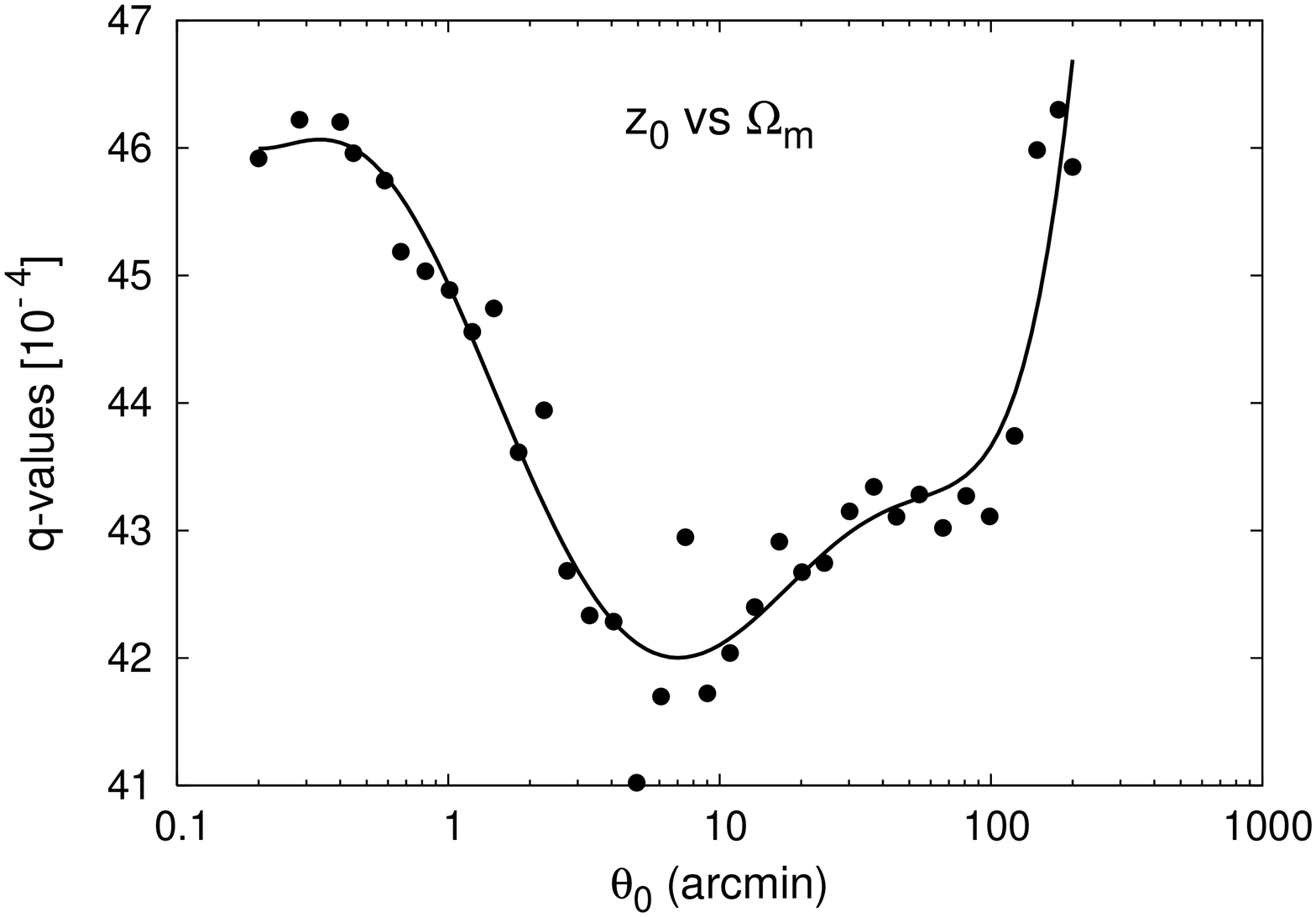}
   \caption{Here we plot $q$ of the combined data vector when varying $(\theta_0)$ of the additional $\xi_+$ data point. We calculate $q$ for 35 different added $\xi_+(\theta_0)$ and show the behaviour in three different parameter spaces. $q$ of the combined data vector can be optimized with respect to $\theta_0$ and the optimal values are $7 \farcm 8$ ($\Gamma$ vs. $\Omega_\mr m$), $12 \farcm 9$ ($\sigma_8$ vs. $\Omega_\mr m$) and $7 \farcm 0$ ($z_0$ vs. $\Omega_\mr m$). These values are the minima of a polynomial fit through the data points.}
         \label{fig:qvar2para}
   \end{figure}

The likelihood analysis in this section is performed in a two-dimensional parameter space; all other cosmological parameters are fixed to the fiducial values. Before comparing the three data vectors we optimize $\vec{\mathcal N}$ with respect to the $\theta_0$-value of the added 2PCF data point. We add 35 different $\xi_+(\theta_0)$ covering a range $\theta_0$ $\in$ [0 \farcm 2-200 \farcm 0] and calculate $q$. Fig. \ref{fig:qvar2para} illustrates the results of this optimization for 3 different pairs of parameters ($\Gamma$ vs. $\Omega_\mr m$, $\sigma_8$ vs. $\Omega_\mr m$, $z_0$ vs. $\Omega_\mr m$).
\begin{figure*}
\sidecaption
   \includegraphics[width=13cm]{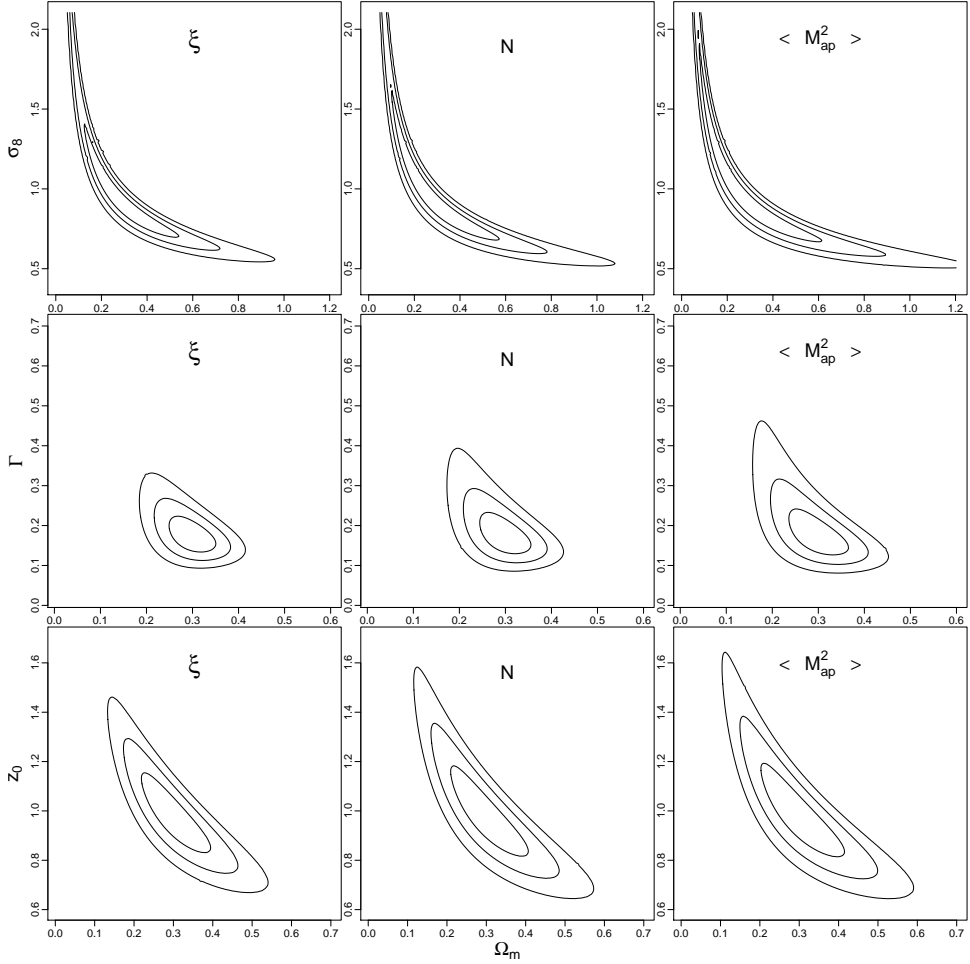}
        \caption{The likelihood contours when varying only two parameters, while the others are fixed to the fiducial values. The contours contain 68.3 \%, 95.4 \%, 99.73 \% of the posterior likelihood. We consider 3 parameter spaces, from top to bottom: $\sigma_8$  vs. $\Omega_\mr m$, $\Gamma$ vs. $\Omega_\mr m$, $z_0$ vs. $\Omega_\mr m$. The constraints of $\vec \xi$ are shown on the left, $\vec{\mathcal M}$ is plotted in the middle and the results of $\vec{\langle M^2_\mr{ap} \rangle}$ are shown on the right.}
         \label{fig:2para}
   \end{figure*}
For all parameter combinations considered the optimal $\theta_0$ is close to $10'$. This can be explained from the behavior of the covariance matrix. For small angular scales the covariance is dominated by shot noise, whereas for large angular scales the signal of $\xi_+$ becomes very small. In both cases the signal-to-noise ratio is lower than at medium angular scales, where we find the minimum of  $q$. In our further analysis we always choose the optimal 2PCF data point for the combined data vector. The results are illustrated by contour plots (Fig. \ref{fig:2para}) and the corresponding values of $q$ are summarized in Table \ref{tab:2paraqvalues}. Here, we also list the results for two additional parameter combinations, $\sigma_8$ vs. $\Gamma$ and $z_0$ vs. $\sigma_8$, not shown in Fig. \ref{fig:2para}.
\begin{table}[h!]
\caption{This table shows the $q$ of $\vec \xi$, $\vec{\mathcal N}$ and $\vec{\langle M^2_\mr{ap} \rangle}$ for various parameter spaces. Parameters over which we marginalize are mentioned in brackets. The entries are given in units of $10^{-4}$ and only $q$ of the same parameter space can be compared. $\Delta \vec{\mathcal N}$ ($\Delta \vec \xi$) gives the relative improvement compared to the $q$ of $\vec{\langle M^2_\mr{ap} \rangle}$ and this improvement differs with respect to the parameter space.}
\begin{center}
\label{tab:2paraqvalues}
\begin{tabular}{|c||c|c|c||c|c|}\hline
parameter space& $\vec{\langle M^2_\mr{ap} \rangle}$&$\vec{\mathcal N}$&$\vec \xi$& $\Delta \vec{\mathcal N}$ & $\Delta \vec{\xi}$  \\ \hline
$\Gamma$ vs. $\Omega_\mr m$&14.7&11.7 &9.1&20.4 \%& 38.1 \%\\ \hline
$\sigma_8$ vs. $\Gamma$ &23.1&19.0 &14.6&17.8 \%&36.8  \%\\ \hline
$\sigma_8$ vs. $\Omega_\mr m$&427.1&314.5&220.1&26.4 \%&48.5 \% \\ \hline
$z_0$ vs. $\Omega_\mr m$ &46.4&41.0&32.9&11.6 \%& 29.1 \%\\ \hline
$z_0$ vs. $\sigma_8$ &95.3&91.4&73.2&4.1 \% &23.2 \%\\ \hline
\hline
$\sigma_8$ vs. $\Omega_\mr m$ ($z_0$)&416.9&313.4&230.0&25.8 \% &44.8 \% \\ \hline
$\sigma_8$ vs. $\Omega_\mr m$ ($\Gamma$)&780.5&720.9&527.0&7.6 \%&32.5 \%\\ \hline
$\Gamma$ vs. $\Omega_\mr m$ ($\sigma_8$)&93.7&77.6&61.6&17.2 \%&34.3 \%\\ \hline
\hline
$\sigma_8$ vs. $\Omega_\mr m$ ($\Gamma$, $z_0$)&983.8&850.6&623.5&13.5 \%&36.6 \%\\ \hline
\end{tabular}
\end{center}
\end{table}
One clearly sees that the 2PCF data vector gives the tightest constraints on cosmological parameters whereas constraints from the aperture mass dispersion are weaker. Although not matching quite the amount of information of $\vec \xi$, the combined data vector is a substantial improvement compared to $\vec{\langle M^2_\mr{ap} \rangle}$. This result is consistent for all parameter combinations we examine; nevertheless the amount of the improvement varies. We calculate the difference in information of $\vec \xi$ and $\vec{\mathcal N}$ relative to $\vec{\langle M^2_\mr{ap} \rangle}$ and denote these values $\Delta \vec \xi$ and $\Delta \vec{\mathcal N}$ (Table \ref{tab:2paraqvalues}). The parameter combination $\sigma_8$ vs. $\Omega_\mr m$ shows a relative improvement of $\Delta \vec{\mathcal N}=26.4 \%$, whereas the improvement is much less for the case $z_0$ vs. $\sigma_8$ ($\Delta \vec{\mathcal N}=4.1 \%$). The amount of new information of $\xi_+ (\theta_0)$ depends on two main issues. First, $\xi_+$ integrates over a very broad range of the power spectrum and it can happen that although $\mathcal P_\mr E$ is sensitive to the parameters considered, the integral over $\mathcal P_\kappa$ is much less. For example, if one varies $\Gamma$, the power spectrum is tilted and looks significantly different, whereas the corresponding $\xi_+(\theta_0)$ might be very similar. Second, $\vec{\langle M^2_\mr{ap} \rangle}$ does not contain information on small Fourier modes, whereas $\vec{\mathcal N}$ gains information about these modes from the data point $\xi_+(\theta_0)$. However, in case these modes of the power spectrum are not sensitive to parameters considered, the information which is contributed by $\xi_+ (\theta_0)$ is mainly redundant, hence $\Delta \vec{\mathcal N}$ is low. For example, varying $\sigma_8$ or $\Omega_\mr m$ changes $\mathcal P_\mr E$ similarly, i.e. increasing $\Omega_\mr m$ or $\sigma_8$ increases the amplitude of $\mathcal P_\mr E$ on all Fourier modes. Therefore, the integration over $\mathcal P_\mr E$ is equally sensitive to parameter variations as $\mathcal P_\mr E$ itself. Furthermore the deviation of power spectra with different values in $\sigma_8$ and $\Omega_\mr m$ becomes much more significant for small Fourier modes. Information on these scales is not included in $\vec{\langle M^2_\mr{ap} \rangle}$ but contributed by $\xi_+ (\theta_0)$, resulting in a large $\Delta \vec{\mathcal N}$(26,4 \%). In contrast to this, a variation of $z_0$ changes the power spectum very little, especially on low $\ell$-scales the dependence is weak. Accordingly, the gain in information for the cases $z_0$ vs. $\Omega_\mr m$ and $z_0$ vs. $\sigma_8$ is rather small.

\subsection{Variation of three and four parameters - marginalization}
\label{sec:3para}
In this section we perform a likelihood analysis in three- and four-dimensional parameter space. To illustrate the results in two-dimensional contour plots we define the marginalized posterior likelihood 
\be
P_\mr{mPL}(\vec \pi_{12}|\vec \xi_\pm)=\int d \pi_3 \int d \pi_4 \, P_\mr{PL}(\vec \pi_{1234}|\vec \xi_\pm) \,,
\ee
which is obtained by integrating over the posterior likelihood of the marginalized parameters. The marginalized likelihood is also biased due to its non-linearity in the log-likelihood. To examine whether this bias affects our results significantly we perform the same experiment as done for $q$ in two-dimensional parameter space.
\begin{figure}[h!]
   \centering
   \includegraphics[width=6cm, angle=270]{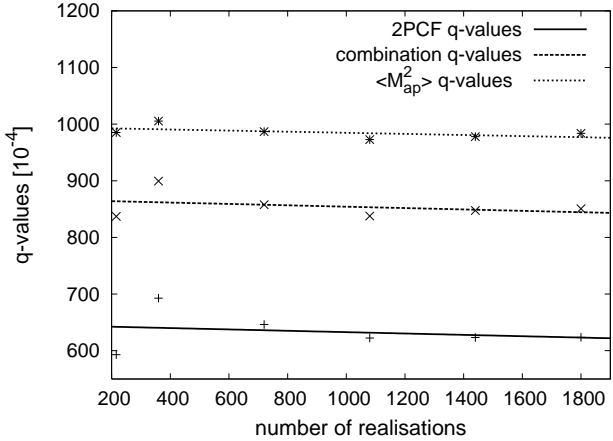}
      \caption{This figure shows the $q$ of $\vec \xi$, $\vec{\mathcal N}$ and $\vec{\langle M^2_\mr{ap} \rangle}$ for the marginalized posterior likelihood depending on different numbers of realisations. The parameter space is $\sigma_8$ vs. $\Omega_\mr m$ (marginalized over $\Gamma$ and $z_0$). The deviation of $q$ belonging to different numbers of realisations is much smaller compared to the deviation of $q$ of different measures. The lines indicate the fit through the data points.}
         \label{fig:qmargvalues}
   \end{figure}
\begin{table}[h!]
\caption{The optimal angular separation $\theta_0$ for the added $\xi_+$ in the combined data vector $\vec{\mathcal N}$. The values are comparable to the similar analysis in two-parameter space (see Fig. \ref{fig:qvar2para}).}
\begin{center}
\label{tab:qvar3para}
\begin{tabular}{|c|c|}\hline
parameter space& optimal value $\theta_0$  \\ \hline
$\Gamma$ vs. $\Omega_\mr m$ (marginalized over $\sigma_8$) &$\theta_0= 9'.1$ \\ \hline
$\sigma_8$ vs. $\Omega_\mr m$ (marginalized over $z_0$) & $\theta_0= 13'.0$ \\ \hline
$\sigma_8$ vs. $\Omega_\mr m$ (marginalized over $\Gamma$ and $z_0$) & $\theta_0=12'.0$ \\ \hline
\end{tabular}
\end{center}
\end{table}
We calculate $q$ for our three different measures depending on the number of realisations. The results are shown in Fig. \ref{fig:qmargvalues}; again, the bias due to the process of marginalization is small compared to the difference of $q$ of our three data vectors showing that also in the marginalized case we can use $q$ to compare the information content.
We also optimize the combined data vector, similar to Sect. \ref{sec:2para} and summarize the results in Table \ref{tab:qvar3para}. 
\begin{figure*}
\sidecaption
   \includegraphics[width=13cm]{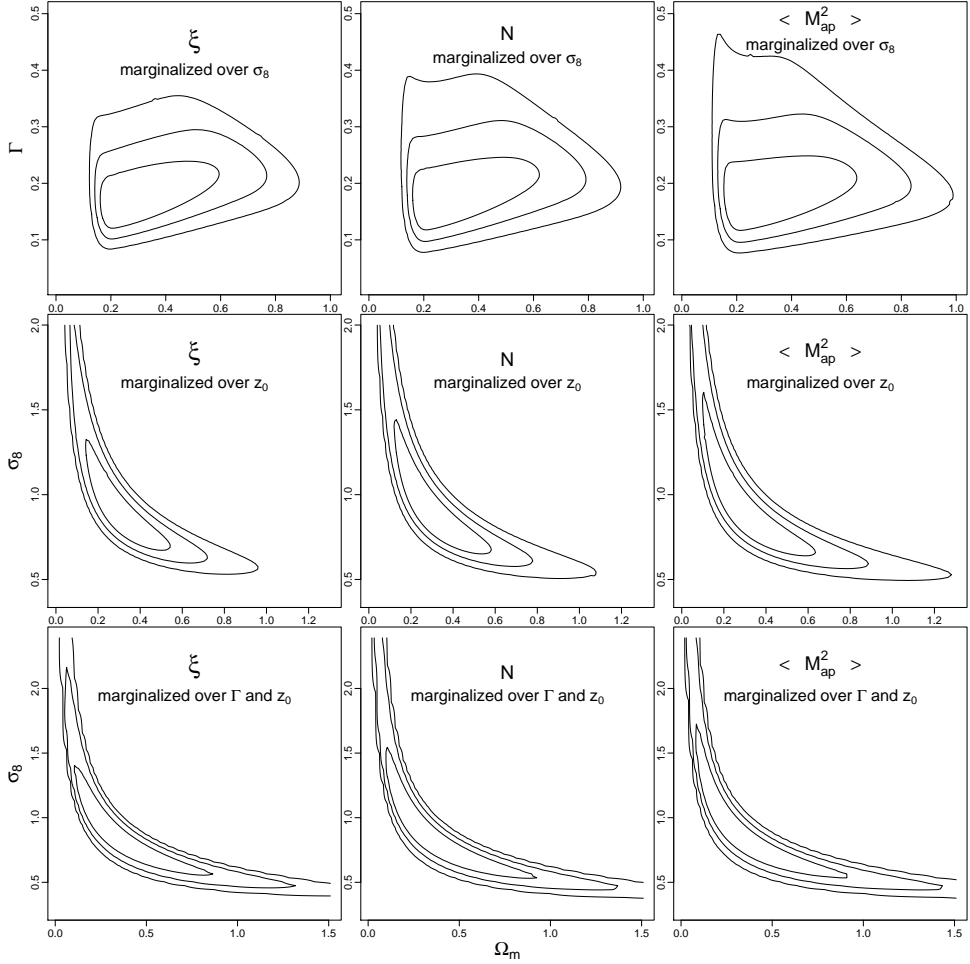}
      \caption{The likelihood contours of $\vec \xi$, $\vec{\mathcal N}$ and $\vec{\langle M^2_\mr{ap} \rangle}$ in three- and four-dimensional parameter space. From top to bottom we see $\Gamma$  vs. $\Omega_\mr m$ marginalized over $\sigma_8$, $\sigma_8$ vs. $\Omega_\mr m$ marginalized over $z_0$ and $\sigma_8$ vs. $\Omega_\mr m$ marginalized over $\Gamma$ and $z_0$. The contours contain 68.3 \%, 95.4 \%, 99.73 \% of the marginalized posterior likelihood. The small scatter of the contours in the last plot is due to a lower resolution of the grid in four-dimensional parameter space compared to the grids in two- and three-dimensional parameter space. The contours, although broader, are comparable to those given in Fig. \ref{fig:2para}.}
         \label{fig:3para}
   \end{figure*}
For the same reasons as in the previous section the optimal angular scale of the added $\xi_+$ data point is again around $10'$ and we choose this optimized $\vec{\mathcal N}$ for the likelihood analysis in three- and four-dimensional parameter space. The results of the likelihood analysis are comparable to those obtained in two-dimensional parameter space. The $q$ (see Table \ref{tab:2paraqvalues}) are larger and the contours (see Fig. \ref{fig:3para}) are broader. 
 Again, the relative improvement $\Delta \vec{\mathcal N}$ depends on the parameter space considered. For $\sigma_8$ vs. $\Omega_\mr m$ marginalized over $z_0$ the improvement is very high (25.8 \%) but becomes much lower for $\sigma_8$ vs. $\Omega_\mr m$ marginalized over $\Gamma$. This can be explained by looking how $\mathcal P_\mr E$ changes with respect to the variation in parameter space. For the combination $\sigma_8$ vs. $\Omega_\mr m$, we already explained this in Sect. \ref{sec:2para} and the influence of $z_0$ on $\mathcal P_\mr E$ is quite similar. Increasing $z_0$ also increases $\mathcal P_\mr E$, although the effect is not very large. Therefore, the improvement of $\sigma_8$ vs. $\Omega_\mr m$ marginalized over $z_0$ is comparable to the non-marginalized case. When varying the shape parameter $\Gamma$,  $\mathcal P_\mr E$ is tilted and this dependence of $\mathcal P_\mr E$ on $\Gamma$ is different compared to the other three parameters. Scales of $\mathcal P_\mr E$ which are most sensitive to $\Gamma$ differ from scales sensitive to $\sigma_8$, $\Omega_\mr m$ and $z_0$ and the same argument holds for the scales of the added $\xi_+(\theta_0)$. Therefore, the optimal $\theta_0 $ for the case $\sigma_8$ vs. $\Omega_\mr m$ marginalized over $\Gamma$ is a compromise and the relative improvement is much lower (7.6 \%) compared to $\sigma_8$ vs. $\Omega_\mr m$ marginalized over $z_0$ (25.8 \%).


\section{Simulation of a B-mode contamination on small angular scales}
\label{sec:Bmodes}
\begin{figure*}
  \includegraphics[width=17cm]{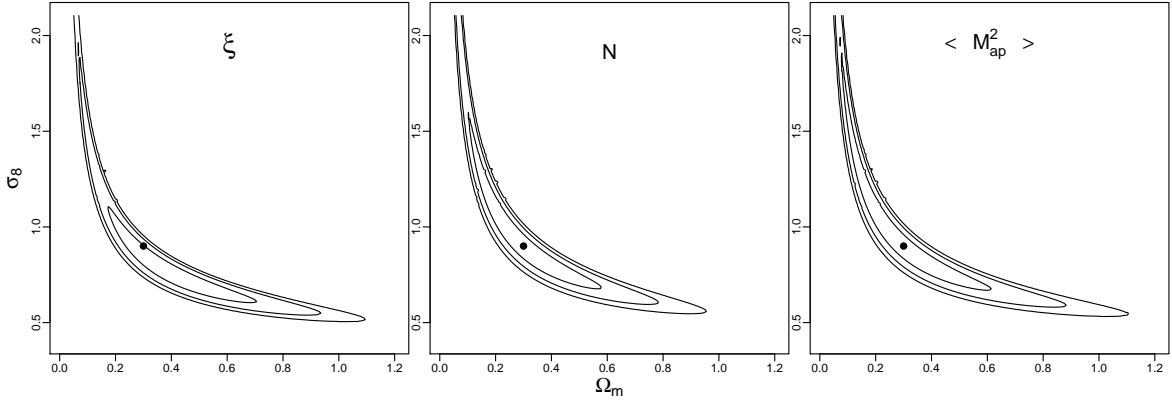}
      \caption{This plot shows the likelihood contours for the case that the shear signal is contaminated with B-modes. We only consider a two-dimensional parameter space ($\sigma_8$ vs. $\Omega_\mr m$) and the contours again contain 68.3 \%, 95.4 \%, 99.73 \% of the posterior likelihood. The black dot in each plot indicates the fiducial model. $\vec \xi$ gives biased constraints, $\vec{\mathcal N}$ and $\vec{\langle M_{\mr{ap}}^2 \rangle}$ are hardly contaminated.}
         \label{fig:b-modes}
   \end{figure*}

In this section we simulate a B-mode contamination of $\vec \xi$, $\vec{\mathcal N}$ and $\vec{\langle M^2_\mr{ap} \rangle}$ on small angular scales. At present there is no model available which describes B-modes; taking into account that B-modes most likely occur on small angular scales \citep[e.g.][]{hoe02,vwmh05,mas07} we use the following arbitrary model for a B-mode power spectrum
\be
\mathcal P_\mr B (\ell)=0.2\, \mathcal P_\mr E (\ell) \; \mr e^{ -\ell_\mr B /\ell} \,,
\ee
where $\ell_\mr B$ defines a scale beyond which the B-mode contamination decreases quickly.
The B-mode contribution to $\vec \xi$ can be calculated from (\ref{eq:xi+}) and (\ref{eq:xi-}) by assuming $\mathcal P_\mr E =0$. In order to calculate the covariance $\mathbf C_\mr B$ we assume that the probability distribution of B-modes can be described by a Gaussian random field. This assumption enables us to calculate the covariance directly in terms of the power spectrum $\mathcal P_\mr B$ \citep{jse07}. The covariance of the 2PCF corresponding to the B-mode contribution is given by
\beq
\tn C^{++}_{\mr B,ij} &=& \frac{1}{\mr A \pi} \int \tn d \ell \ell  \mr J_0 \left( \ell \vartheta_i \right) J_0 \left(\ell \vartheta_j \right)  \left( \mathcal P^2_\mr B(\ell)+ \mathcal P_\mr B(\ell) \frac{\sigma^2_\epsilon}{n} \right) \nn \,, \\
\tn C^{--}_{\mr B,ij} &=& \frac{1}{\mr A \pi} \int \tn d \ell \ell  \mr J_4 \left( \ell \vartheta_i \right) J_4 \left(\ell \vartheta_j \right)  \left( \mathcal P^2_\mr B(\ell)+ \mathcal P_\mr B(\ell) \frac{\sigma^2_\epsilon}{n} \right) \nn \,, \\
\tn C^{+-}_{\mr B,ij} &=& - \frac{1}{\mr A \pi} \int \tn d \ell \ell  \mr J_0 \left( \ell \vartheta_i \right) J_4 \left(\ell \vartheta_j \right)  \left( \mathcal P^2_\mr B(\ell)+ \mathcal P_\mr B(\ell) \frac{\sigma^2_\epsilon}{n} \right) \nn\,,
\eeq
where $\mr A$ defines the volume of the survey, $\sigma_\epsilon$ the intrinsic ellipticity noise and $n$ the number density of the source galaxies. According to the corresponding values of the ray-tracing simulations we choose $\sigma_\epsilon =0.3$ and $n=25/\tn{arcmin}^2$. Note that $\tn C^{-+}_{\mr B,ij}=\tn C^{+-}_{\mr B,ji}$. The pure shot noise term of $\mathbf C^{\pm \pm}_{\mr B}$ is contained in $\mathbf C^{\pm \pm}_{\mr E}$, in case of $\mathbf C^{+-}_{\mr B}$ this term vanishes anyway. We further assume that the contamination is independent of the lensing signal, hence there is no correlation between E- and B-modes. This assumption does not hold in case the B-mode signal is caused by insufficient PSF correction or other systematics, and we will comment on this at the end of this section. For the case that B-modes are created independently from E-modes we can define a combined E/B-mode covariance matrix as
\be
\mathbf C_\mr{tot}= \mathbf C_\mr E + \mathbf C_\mr B\,.
\ee
Recall that $\bfmath C_\mr E$ is estimated from ray-tracing simulations whereas $\bfmath C_\mr B$ is calculated by assuming a Gaussian random field. The correction factor, needed to invert estimated matrices correctly (see Sect. \ref{sec:mixing}), must only be applied to  $\bfmath C_\mr E$, not for  $\bfmath C_\mr B$. We use the iterative approach of \cite{ken81} to decompose this inverse of a sum of matrices into a summation of inverse matrices and then apply the correction factor only to $\mathbf C_\mr E^{-1}$. From now on, the procedure of the comparison is similar to Sect. \ref{sec:2para} and Sect. \ref{sec:3para}. We calculate the $\mathbf C_\mathcal M$ and $\mathbf C_\mathcal N$ from $\bfmath C_\xi$ and perform a likelihood analysis. We only show the results for the $\Omega_\mr m$ vs. $\sigma_8$ plane (see Fig. \ref{fig:b-modes}).
 The black dots indicate the fiducial cosmological model, and in case of the 2PCF data vector there is a significant deviation to the parameters of the maximum of the posterior likelihood. $\vec{\langle M_{\mr{ap}}^2 \rangle}$ and $\vec{\mathcal N}$ are much more robust against the contamination. As expected, the maximum of the posterior likelihood of the aperture mass dispersion matches exactly the fiducial parameters and in case of $\vec{\mathcal N}$ the discrepancy is negligibly small. Furthermore, the combination still gives tighter constraints on the parameters. As already mentioned above, the assumption of B-modes being independent of the E-mode signal does not always hold. In case the contamination affects both, E-mode and B-mode signal, the impact on the parameter constraints of the different measures is hard to quantify. In case one measures a B-modes signal, it is a common approach to assume that the E-mode signal is contaminated in a similar way, hence one correspondingly increases its error bars. Although this assumption is sensible, there are possible scenarios where the amount of contamination in E- and B-mode differs and the E-mode contamination cannot be quantified at all. Under the assumption that B-modes trace the scales of the E-mode contamination it is reasonable to exclude those scales from the likelihood analysis. This can be done using $\vec{\langle M_{\mr{ap}}^2 \rangle}$ or $\vec{\mathcal N}$ but $\vec \xi$ cannot avoid the contamination due to its broad filter functions.

\section{Conclusions}
\label{sec:conc}
Although the 2PCF and the aperture mass dispersion are both filtered versions of the power spectrum the first contains more information on $\mathcal P_\mr E$ than the latter. Reason for this is that $\vec \xi$ samples the power spectrum over a much broader range and also collects information on scales which are larger than the size of the survey. $\vec{\langle M_{\mr{ap}}^2 \rangle}$ lacks this large-scale information, but yields highly localized information on $\mathcal P_\mr E$. Nevertheless $\vec{\langle M_{\mr{ap}}^2 \rangle}$ has other advantages. First, due to its narrow filter function the data points are much less correlated compared to the 2PCF data points. This leads to a mainly diagonal covariance matrix, which is numerically more stable during the inversion process in a likelihood analysis. Second, when considering higher-order statistics ${\langle M^3_\mr{ap} \rangle}$ is much easier to handle than the three-point correlation function \citep{skl05} and third, the aperture mass dispersion is sensitive to E-modes only. Based on these considerations we create the combined data vector $\vec{\mathcal N}$, which preserves the advantages of $\vec{\langle M_{\mr{ap}}^2 \rangle}$ and additionally provides large-scale information on $\mathcal P_\mr E$. This data vector can be optimized with respect to the angular scale of the added data point $\xi_+(\theta_0)$, but this optimization very likely depends on the survey geometry and must be performed for each survey separately. We compare the three data vectors in a detailed likelihood analysis and find that the combined data vector is a strong improvement in information content compared to $\vec{\langle M_{\mr{ap}}^2 \rangle}$. However, the amout of improvement depends on the parameter space considered, more precisely, on the dependence of $\mathcal P_\mr E$ on variation of those parameters. The combined data vector $\vec{\mathcal N}$ also maintains the other advantages of the aperture mass dispersion. Its covariance matrix is almost diagonal, even the cross terms $\tn{C}(\mathcal M(\theta_k),\hat \xi_+(\theta_0))$ are much smaller compared with the off-diagonal terms of $\bfmath C_\xi$. Comparing the information content of $\vec \xi$ and $\vec{\mathcal N}$,  $\vec \xi$ gives tighter constraints if the shear signal only consists of E-modes. In the more realistic case, when also B-modes are present, $\vec \xi$ is biased whereas $\vec{\mathcal N}$ is hardly affected and still gives tighter constraints on cosmological parameters compared to $\vec{\langle M_{\mr{ap}}^2 \rangle}$.
\begin{appendix}

\section{Comparison of two measures}
\label{sec:general}
We compare the information content of two arbitrary data vectors referring to them as primary data vector $\vec p$ and secondary data vector $\vec s$. We further assume that $\vec s$ can be calculated from $\vec p$ by a \textit{transfer matrix} $\bfmath A$ (dimension $n \times m$), with arbitrary $n$ and $m$
\be
\label{eq:assump}
\vec p = \left( \begin{array}{c}
		p_1  \\
		p_2 \\
		\vdots \\
		p_m
		\end{array}\right) \; \quad \tn{and} \quad
\vec s = \left( \begin{array}{c}
		s_1  \\
		s_2 \\
		\vdots \\
		s_n
		\end{array} \right) \; \quad \tn{with} \;\quad \vec s= \mathbf A \; \vec p\;.
\ee
We define the covariance matrices of these data vectors as 
\beq
\label{eq:covp}
\mathbf C_\mr p&=&\left \langle (\vec p -\vec{\hat p}) (\vec p - \vec{\hat p})^\mr t \right \rangle \,,\\
\mathbf C_\mr s&=&\left \langle (\vec s -\vec{\hat s}) (\vec s - \vec{\hat s})^\mr t \right \rangle \,,
\eeq
where $\vec{\hat p}$ ($\vec{\hat s}$) denotes the estimated and $\vec p$ ($\vec s$) the true values of primary (secondary) measure. Using (\ref{eq:assump}) we can relate both covariances through
\be
\label{eq:assump2}
\mathbf C_\mr s= \mathbf A \; \mathbf C_\mr p \;\mathbf  A^\mr t \;.
\ee
The transformation matrix $\mathbf A$ has to be of rank $\mathbf A=n$, otherwise the covariance matrix of the secondary data vector $\mathbf C_\mr s=(\mathbf A \, \mathbf C_\mr p \, \mathbf A^\mr t)$ is singular and not invertible. Furthermore as $\mathbf A$ is of dimension ($n \times m$), rank $\mathbf A \leq m$ implying $n \leq m$. We take the $\chi^2$-functions a measure for the information content 
\be
\label{eq:chiprimsec}
\chi^2_\mr p= \vec \Delta_\mr p^\mr t \; \mathbf C_\mr p^{-1} \; \vec \Delta_\mr p  \; \qquad \tn{and} \qquad \chi^2_\mr s = \vec \Delta_\mr s^\mr t \;\mathbf  C_\mr s^{-1} \;\vec \Delta_\mr s \; ,
\ee
where in our case $\vec \Delta_\mr p = \vec p^{\mr f} - \vec p_{\vec \pi}$ ($\vec \Delta_\mr s = \vec s^{\mr f} - \vec s_{\vec \pi}$) denotes the difference between the fiducial data vector $\vec p^{\mr f}$ ($\vec s^{\mr f}$) and the data vector $\vec p_{\vec \pi}$ ($\vec s_{\vec \pi}$) depending on the parameter vector $\vec \pi$. In case $\chi^2$ is minimal, the posterior likelihood of the corresponding $\vec \pi$ being the correct parameter vector is maximized. The difference between $\chi^2_\mr p$ and $\chi^2_\mr s$ characterizes which probability function has a larger curvature, i.e. which data vector gives tighter constraints in parameter space. Therefore the information content of primary and secondary data vector can be compared by calculating 
\be
\label{eq:conjecoriggeneral}
\chi^2_\mr p \,-\,\chi^2_\mr s = \vec \Delta_\mr p^\mr t \; \mathbf C_\mr p^{-1} \; \vec \Delta_\mr p \; - \; \vec \Delta_\mr p^\mr t \; \mathbf A^\mr t \; \left( \mathbf A \; \mathbf C_\mr p \;\mathbf A^\mr t \right)^{-1} \mathbf{A} \; \vec \Delta_\mr p \;,
\ee
for arbitrary $\vec \Delta_\mr p$. In case this difference is always positive we can conclude that the primary data vector gives tighter constraints on parameters. We can always find transformation matrices $\mathbf V$ (dimension $m \times m$) and $\mathbf U$ (dimension $n \times n$) to rewrite the transfer matrix $\mathbf A$ as an $n \times m$ matrix 
\be
\label{eq:trafo1}
\left( \begin{array}{cccccccc}
		\multicolumn{7}{c|}{\rule[-3mm]{0mm}{8mm}\textbf{$\bfmath E_n$}}& \multicolumn{1}{c}{\textbf{$\bfmath 0$}}
 \end{array} \right) \;= \; \bfmath S \; = \bfmath U \; \mathbf A \; \mathbf V^{-1} \quad \longleftrightarrow \quad \bfmath A\; = \; \bfmath U^{-1} \; \mathbf S \; \mathbf V \,.
\ee
We can directly calculate these transformation matrices as a multiplication of elementary matrices \citep{fitrafo}. Inserting (\ref{eq:trafo1}) into (\ref{eq:conjecoriggeneral}) we derive after some lengthy but straightforward calculation 
\be
\label{eq:conjecfin}
\chi^2_\mr p \; - \; \chi^2_\mr s = \vec \Delta_\mr p'^{\mr t} \; \mathbf C'^{-1} \; \vec \Delta_\mr p'\; - \; \vec \Delta_\mr p'^{\mr t} \; \mathbf S^\mr t \; \left( \mathbf S \;\mathbf C' \; \mathbf S^\mr t \right)^{-1} \mathbf S \; \vec \Delta_\mr p'
\ee
with 
\be
\label{eq:Candpnew}
\mathbf C' = \mathbf V \; \mathbf C_\mr p \; \mathbf V^\mr t \qquad \tn{and} \qquad  \vec \Delta_\mr p'=\mathbf V \; \vec \Delta_\mr p \,.
\ee
For simpler notation we discard all  `` $'$ " further on. 
We define 
\be 
\label{eq:Cinvers_p}
\mathbf C^{-1}= \left( \begin{array}{cccccccc}
		\multicolumn{7}{c|}{\rule[-3mm]{0mm}{8mm}\textbf{$\bfmath C_1$}}& \multicolumn{1}{c}{\textbf{$\bfmath C_2$}}\\
		\hline
		\multicolumn{7}{c|}{\rule[1mm]{0mm}{3mm}\textbf{$\bfmath C_2^\mr t$}}& \multicolumn{1}{c}{\textbf{$\bfmath C_3$}}
		 \end{array} \right)^{-1} = \left( \begin{array}{cccccccc}
		\multicolumn{7}{c|}{\rule[-3mm]{0mm}{8mm}\textbf{$\bfmath D_1$}}& \multicolumn{1}{c}{\textbf{$\bfmath D_2$}}\\
		\hline
		\multicolumn{7}{c|}{\rule[1mm]{0mm}{3mm}\textbf{$\bfmath D_2^\mr t$}}& \multicolumn{1}{c}{\textbf{$\bfmath D_3$}}
		 \end{array} \right),
\ee
with $\mathbf C_1$ being an $n \times n$ matrix and calculate 
\be
\label{eq:Cinvers_s}
\mathbf S^\mr t \; \left( \mathbf S \;\mathbf C \; \mathbf S^\mr t \right)^{-1} \mathbf S = \left( \begin{array}{cccccccc}
		\multicolumn{7}{c|}{\rule[-3mm]{0mm}{8mm}\textbf{$\bfmath C_1^{-1}$}}& \multicolumn{1}{c}{\textbf{$\bfmath 0$}}\\
		\hline
		\multicolumn{7}{c|}{\rule[1mm]{0mm}{3mm}\textbf{$\bfmath 0$}}& \multicolumn{1}{c}{\textbf{$\bfmath 0$}}
		 \end{array} \right) \,.
\ee
Using (\ref{eq:Cinvers_p}) and (\ref{eq:Cinvers_s}) we can rewrite (\ref{eq:conjecfin}) as 
\be
\label{eq:d1-c1}
\chi^2_\mr p \; -\; \chi^2_\mr s   = \vec \Delta_\mr p^\mr t \; \left( \begin{array}{cccccccc}
		\multicolumn{7}{c|}{\rule[-3mm]{0mm}{8mm}\textbf{$\bfmath D_1-\bfmath C_1^{-1}$}}& \multicolumn{1}{c}{\textbf{$\bfmath D_2$}}\\
		\hline
		\multicolumn{7}{c|}{\rule[1mm]{0mm}{3mm}\textbf{$\bfmath D_2^\mr t$}}& \multicolumn{1}{c}{\textbf{$\bfmath D_3$}}
		 \end{array} \right) \; \vec \Delta_\mr p \,.
\ee
From $\mathbf C \, \mathbf D = \mathbf E_m$ we deduce 
\be
\label{eq:sub1}
\bfmath C_1 \; \mathbf D_1 + \mathbf C_2 \;\mathbf D_2^\mr t  = \bfmath E_n \; \longrightarrow \; \bfmath D_1 - \mathbf C_1^{-1} = - \mathbf C_1^{-1} \; \mathbf C_2 \; \mathbf D_2^\mr t 
\ee
and
\be
\label{eq:sub2}
\bfmath C_1 \; \mathbf D_2 + \mathbf C_2 \; \mathbf D_3  = 0 \; \longrightarrow \; \bfmath C_2 = - \bfmath C_1 \; \mathbf D_2 \; \mathbf D_3^{-1} \,.
\ee 
Inserting (\ref{eq:sub2}) into (\ref{eq:sub1}) we can rewrite (\ref{eq:d1-c1}) as
\be
 \chi^2_\mr p - \chi^2_\mr s= \vec \Delta_\mr p^\mr t \; \left( \begin{array}{cccccccc}
		\multicolumn{7}{c|}{\rule[-3mm]{0mm}{8mm}\textbf{$\bfmath D_2 \bfmath D_3^{-1} \bfmath D_2^\mr t$}}& \multicolumn{1}{c}{\textbf{$\bfmath D_2$}}\\
		\hline
		\multicolumn{7}{c|}{\rule[1mm]{0mm}{3mm}\textbf{$\bfmath D_2^\mr t$}}& \multicolumn{1}{c}{\textbf{$\bfmath D_3$}}
		 \end{array} \right)
		 \; \vec \Delta_\mr p \,.
\ee
$\bfmath C$ is positive definite and symmetric, therefore $\bfmath D_3$ as a submatrix is positive definite and symmetric and also the inverse $\bfmath D_3^{-1}$ has these favorable properties \citep{and}. Hence, we can decompose $\mathbf D_3=\mathbf L \mathbf L^\mr t$ and finish our calculation as follows
\beq
\chi^2_{\mr p} - \chi^2_{\mr s}&=& \vec \Delta_\mr p^\mr t \; \left( \begin{array}{cccc}
		\multicolumn{4}{c}{\rule[-3mm]{0mm}{8mm}\textbf{$\bfmath D_2 (\bfmath L^\mr t)^{-1}$}}\\
		\multicolumn{4}{c}{\rule[1mm]{0mm}{3mm}\textbf{$\bfmath L$}}
		\end{array} \right)\;
		\underbrace{\left( \bfmath L^{-1} \; \bfmath D_2^\mr t \;\; \bfmath L \right)}_{\mathbf T} \; \vec \Delta_\mr p  \\ \nonumber
&=& \vec \Delta_\mr p^\mr t \bfmath T^\mr t \, \bfmath T  \vec \Delta_\mr p \nonumber \\ 
&=& || \bfmath T  \vec \Delta_\mr p ||^2 \nonumber\\
&\geq& 0.
\eeq
We will now examine the case when $\chi^2_\mr p - \chi^2_\mr s = 0$. The information content of primary and secondary measure is considered to be equal if and only if this equality holds for \textbf{all} data vectors $\vec \Delta_\mr p$. If there is only one $\vec \Delta_\mr p$ for which $\chi^2_\mr p- \chi^2_\mr s > 0$, the primary measure contains more information. The difference of the two $\chi^2$-values is given by (\ref{eq:conjecoriggeneral}). In case it is zero for all $\vec \Delta_\mr p$, 
\be
\label{eq:zerocase}
 \mathbf C_\mr p^{-1} =  \mathbf A^\mr t \; \left( \mathbf A \; \mathbf C_\mr p \;\mathbf A^\mr t \right)^{-1} \mathbf{A}
\ee 
must hold \citep{fizero}. $\mathbf C_\mr p$ is of rank $m$, hence the lefthandside of (\ref{eq:zerocase}) must also have rank $m$. Then $\mathbf A$ must have rank $m$ and is therefore a quadratic $m \times m$ matrix, which is of course invertible. This result is intuitively clear, if one is able to calculate $\vec \Delta_\mr s$ from $\vec \Delta_\mr p$ and vice versa the information content should be the same.
We can summarize the results of the above calculation in two statements:
\begin{enumerate}
\item If a secondary measure can be calculated from a primary by a matrix $\bfmath A$ as described in (\ref{eq:assump}), the secondary measure has less or equal information.
\item The amount of information is equal in case the rank of $\bfmath A$ equals the dimension of the primary data vector ($m$) implying that $\bfmath A$ is invertible.
\end{enumerate}
\end{appendix}

\bibliographystyle{aa}


\begin{acknowledgements}
We thank Yannick Mellier, Jan Hartlap and Tim Schrabback for useful discussions and advise. This work was supported by the Deutsche Forschungsgemeinschaft under the projects SCHN 342/6--1 and SCHN 342/9--1. TE is supported by the International Max-Planck Research School at the University Bonn. MK is supported by the CNRS ANR ``ECOSSTAT'', contract number ANR-05-BLAN-0283-04. 

\end{acknowledgements}

\end{document}